\documentclass[letterpaper,twocolumn,10pt,hyphens]{article}
\usepackage{usenix}

\usepackage{tikz}
\usepackage{dirtytalk}
\usepackage{amsmath}
\usepackage{booktabs}
\usepackage{multirow}
\usepackage{microtype}
\usepackage{hyperref}
\usepackage{cleveref}
\usepackage{todonotes}
\usepackage[inline]{enumitem}
\usepackage{forest}
\usepackage[dvipsnames]{xcolor}
\usepackage{adjustbox}
\usepackage{subcaption}
\usetikzlibrary{arrows,calc,fit,matrix,positioning,shapes,arrows.meta,backgrounds}
\usepackage{fnpct}
\usepackage{listings}
\usepackage{tcolorbox}
\newfloat{lstfloat}{htbp}{lop}
\floatname{lstfloat}{Listing}
 \crefalias{lstfloat}{listing}
\makeatletter
\pgfdeclareshape{datastore}{
  \inheritsavedanchors[from=rectangle]
  \inheritanchorborder[from=rectangle]
  \inheritanchor[from=rectangle]{center}
  \inheritanchor[from=rectangle]{base}
  \inheritanchor[from=rectangle]{north}
  \inheritanchor[from=rectangle]{north east}
  \inheritanchor[from=rectangle]{east}
  \inheritanchor[from=rectangle]{south east}
  \inheritanchor[from=rectangle]{south}
  \inheritanchor[from=rectangle]{south west}
  \inheritanchor[from=rectangle]{west}
  \inheritanchor[from=rectangle]{north west}
  \backgroundpath{
    \southwest \pgf@xa=\pgf@x \pgf@ya=\pgf@y \northeast \pgf@xb=\pgf@x \pgf@yb=\pgf@y \pgfpathmoveto{\pgfpoint{\pgf@xa}{\pgf@ya}}
    \pgfpathlineto{\pgfpoint{\pgf@xb}{\pgf@ya}}
    \pgfpathmoveto{\pgfpoint{\pgf@xa}{\pgf@yb}}
    \pgfpathlineto{\pgfpoint{\pgf@xb}{\pgf@yb}}
 }
}
\makeatother

\tikzset{
	dfdbase/.style={draw,very thick,align=center,font=\ttfamily\footnotesize},
	process/.style={dfdbase,circle, minimum width=1.25cm},
    sprocess/.style={dfdbase,circle, minimum width=.25cm},
	datastore/.style={dfdbase,shape=datastore,inner sep=.1cm,minimum width=1.5cm},
    sdatastore/.style={dfdbase,shape=datastore,inner sep=.1cm,minimum width=.25cm},
	externalentity/.style={dfdbase,inner sep=.1cm,minimum width=1.5cm},
	to/.style={->,>=stealth',shorten >=1pt,very thick,font=\sffamily\footnotesize},
    toway/.style={<->,>=stealth',shorten >=1pt,shorten <=1pt,very thick,font=\sffamily\footnotesize},
	trustboundary/.style={dfdbase,line width=2pt,rectangle,color=red,dashed},
	trustboundaryline/.style={dfdbase,line width=2pt,color=red,dashed},
	dflabel/.style={font=\ttfamily\footnotesize}
}
\tikzstyle{background}=[rectangle,
fill=gray!10,
inner sep=0.2cm,
rounded corners=5mm] \newcommand{\multilinecaption}[3]{
    \caption{#1#2\\\textnormal{\emph{#3}}}
}
\newlength{\landscapewidth}
\newcommand{\excount}{100}
\newcommand{\casethreatcount}{224}
\newcommand{\linddun}{\textsc{linddun}}

\usepackage{threeparttable}
\usepackage{colortbl}
\usepackage{array}
\newcolumntype{P}[1]{>{\centering\arraybackslash}p{#1}}
\newcolumntype{R}[1]{>{\raggedleft\arraybackslash}p{#1}}
\newcolumntype{L}[1]{>{\raggedright\arraybackslash}p{#1}}

\usepackage{balance}

\newcolumntype{M}[1]{>{\centering\arraybackslash}m{#1}}
\newcolumntype{L}[1]{>{\raggedright\arraybackslash}m{#1}}
\usepackage{multicol}
\usepackage[table]{xcolor}

\definecolor{tagRed}{HTML}{F4A7A7}      \definecolor{tagOrange}{HTML}{F8C291}   \definecolor{tagYellow}{HTML}{F6E58D}   \definecolor{tagGreen}{HTML}{B8E994}    \definecolor{tagTeal}{HTML}{A8DADC}     \definecolor{tagBlue}{HTML}{AFCBFF}     \definecolor{tagPurple}{HTML}{D5B6E8}   \definecolor{tagGray}{HTML}{D3D3D3}     

\newcommand{\datatag}[2]{\colorbox{#1}{\strut\textsf{#2}}}

\newcommand{\UI}{\datatag{tagRed}{User Inputs}}
\newcommand{\DA}{\datatag{tagOrange}{Derived Attributes}}
\newcommand{\FTD}{\datatag{tagYellow}{FT Data}}
\newcommand{\PTD}{\datatag{tagGreen}{PT Data}}
\newcommand{\OD}{\datatag{tagTeal}{Operation Data}}
\newcommand{\UD}{\datatag{tagBlue}{User Data}}
\newcommand{\SP}{\datatag{tagPurple}{System Prompts}}

\usepackage{calc}

\newlength{\barW}
\definecolor{barbg}{HTML}{EEEEEE}
\definecolor{barfg}{HTML}{8FBCE6} 

\newcommand{\pctbar}[1]{\begingroup
  \setlength{\fboxsep}{0pt}\colorbox{barbg}{\makebox[\barW][l]{\strut
      \colorbox{barfg}{\rule{\dimexpr \barW*#1/100\relax}{2ex}}\hspace{2pt}\footnotesize #1\% }}\endgroup
}

\begin{document}

\date{}

\title{\Large \bf A \linddun{}-based Privacy Threat Modeling Framework for GenAI}

\def\plainauthor{Liao et al.}

\author{
{\rm Qianying Liao$^*$, Jonah Bellemans$^*$, Laurens Sion}\\
DistriNet, KU Leuven
\and
{\rm Xue Jiang, Dmitrii Usynin, Xuebing Zhou}\\
Huawei Heisenberg Research Center (Munich)
\and
{\rm Dimitri Van Landuyt}\\
LIRIS, KU Leuven
\and
{\rm Lieven Desmet, Wouter Joosen}\\
DistriNet, KU Leuven
} 

\maketitle
\def\thefootnote{*}\footnotetext{These authors contributed equally to this work.}\def\thefootnote{\arabic{footnote}}

\begin{abstract}

As generative AI (GenAI) systems become increasingly prevalent across various technological stacks, the question of how such systems handle sensitive and personal data flows becomes increasingly important.
Specifically, both the ability to harness and process large swaths of information as well as their stochastic nature raise key concerns related to both security and privacy.
Unfortunately, while some of the traditional security threat modeling can effectively identify certain violations, privacy-related issues are often overlooked.
\\
To respond to these challenges, we introduce a novel domain-specific privacy threat modeling framework to support the privacy threat analysis of GenAI-based applications. This framework is constructed through a two-pronged approach: (1)~a systematic review of the emerging literature on GenAI privacy threats, and (2)~a case-driven application to a representative Chatbot system.
These efforts yield a foundational GenAI privacy threat modeling framework built on \linddun{}.
The new framework affects three out of the seven privacy threat types of \linddun{} and introduces \excount{} new GenAI examples to the knowledge base. 
Its effectiveness is validated on an AI Agent system, which demonstrates that a comprehensive privacy analysis can be supported by the new framework.

\end{abstract}

\section{Introduction}
Generative AI (`GenAI') technologies have emerged to become a pervasive component in modern software.
While Large Language Models (LLMs) are among the most prominent examples, GenAI also includes image, audio, video, and code generation systems. 
These technologies automate tasks, assist users, and process information intuitively, lowering adoption barriers. 
As a result, they are increasingly embedded in applications that interact with users and other GenAI-based agents to perform complex tasks.

However, the integration of these technologies into applications poses substantial security and privacy risks~\cite{shanmugarasa2025sok, hong2025sok, ma2025sok, smith2023identifying}.
This is well-illustrated by the growing collection of novel security and privacy attacks targeting the models used for GenAI in research~\cite{shanmugarasa2025sok, hong2025sok, ma2025sok, smith2023identifying}, but also in real-world incidents~\cite{incidents}.
A recent study discovers that ``even the most capable models such as GPT-4 and ChatGPT reveal private information in contexts that humans would not, 39\% and 57\% of the time, respectively''~\cite{mireshghallah2023can}.
This raises concerns about the potential risks of GenAI to reveal sensitive information when incorporated into complex enterprise systems.

Unsurprisingly, the adoption of GenAI introduces societal concerns~\cite{kelley2023,Klymenko2025}, prompting legislative initiatives such as the EU AI Act~\cite{EU2024}, which require organizations to adopt a risk-based approach to security and privacy. 
Yet, complying with legal obligations does not address the practical challenges organizations face when integrating this technology into their systems and services.
The central challenge is how to systematically identify these concerns within their applications.
Several threat modeling frameworks exist~\cite{Sion2025,kunz2023privacy,deng2011privacy,Katcher2024,Wang2024,Long2020} that organizations can use to systematically analyze their applications. 
However, none of these frameworks is specifically tailored to the privacy threat modeling of GenAI systems.
Consequently, they remain too generic and lack sufficient depth to adequately support software engineers during the privacy threat modeling of specialized GenAI-based applications. 
Applying generic knowledge of privacy threats to such systems is therefore a non-trivial endeavor. 

In this work, we focus on two fundamental research questions: 
\textbf{(RQ1)}~What new privacy threats arise specifically in GenAI-based systems? \textbf{(RQ2)}~How can this knowledge of sophisticated GenAI privacy threats be made accessible to practitioners who lack the necessary expertise?

\paragraph{Contribution statement.}~We present a domain-specific privacy threat modeling framework for Generative AI systems, explicitly designed to be usable by software engineers without deep GenAI expertise. 
The framework combines (i)~a \textbf{bottom-up} privacy threat analysis of a GenAI HR chatbot system and (ii)~a \textbf{top-down} systematization of 58 state-of-the-art (SotA) papers on privacy in GenAI-based systems. 
Our main contribution is a \textbf{practical, engineer-oriented privacy threat modeling framework for GenAI-based systems, built on the \linddun{} privacy threat knowledge base} that translates fragmented research insights into actionable guidance for real-world GenAI development and deployment. 
We validate the framework on a multi-agent GenAI assistant case study. We provide the case studies and the GenAI-specific privacy threat knowledge base as supplementary material~\cite{supplementaryMaterials}.

The remainder of this paper is structured as follows.
\Cref{sec:background} provides the background and related work on privacy threat modeling and GenAI.
Next, \cref{sec:methodology} explains the methodology for constructing knowledge on generative AI privacy threats.
\Cref{sec:threats} synthesizes the privacy threats in the SotA and case studies. \Cref{sec:validation} presents the validation of the framework.
\Cref{sec:threatknowledge} introduces the new GenAI privacy threat modeling framework. 
After that, \cref{sec:ttv} discusses threats to validity.
Finally, \cref{sec:conclusion} concludes the paper.

 \section{Background \& related work}\label{sec:background}

This section provides background and describes related work on threat modeling (\cref{ssec:backptm}) and GenAI (\cref{ssec:backgenai}).

\subsection{Privacy threat modeling methodologies}\label{ssec:backptm}

Privacy threat modeling frameworks~\cite{Sion2025,Katcher2024,Wang2024} comprise a set of approaches to support a systematic and comprehensive analysis to identify and evaluate potential privacy risks based on an architectural model of the system under analysis~\cite{Sion2025}.
Those approaches can consist of both a supporting knowledge base~\cite{Sion2025} or taxonomy~\cite{Katcher2024,Wang2024} (capturing relevant knowledge about potential privacy threats) and a methodology that provides a concrete process for practitioners to follow to identify privacy threats in a specific system model.
Based on such an analysis, relevant mitigating measures can be taken, such as using Privacy-Enhancing Technologies, privacy patterns, and privacy design strategies to improve system design and address privacy threats.
On the other hand, the rapid adoption of generative AI has led to numerous initiatives to assess its security and privacy implications.

\paragraph{Security threat modeling.}
There are many generic approaches that can be leveraged for security threat analysis, such as threat modeling with \textsc{stride}~\cite{Shostack2014,Howard2006,kohnfelder1999threats} using DFDs~\cite{DeMarco1979}, attack trees and decision trees~\cite{Schneier1999,Shortridge2021}, or \textsc{pasta}~\cite{UcedaVelez2015}.
These approaches can be further supported by a wide range of knowledge resources to help security analysts to come up with more specific threats, for example, \textsc{capec}~\cite{CAPEC} which provides attack patterns, \textsc{cwe}~\cite{CWE} for identifying the underlying causes of vulnerabilities~\cite{CVE}, \textsc{cawe}~\cite{santos2017catalog} which abstracts these to common architectural design flaws, and the Top 10 Secure Design Flaws~\cite{Arce2014}.
To automate this analysis, numerous threat modeling tools exist that elicit these threats and integrate one or more of the above threat knowledge resources.
Examples of common threat modeling tools are the Microsoft Threat Modeling Tool~\cite{MicrosoftCorporation2025}, ThreatDragon~\cite{OWASP2025ThreatDragon}, pyTM~\cite{Tarandach2025}, \textsc{ovvl}~\cite{Schaad2019}, threagile~\cite{Threagile2025}, IriusRisk~\cite{IriusRisk}, threats manager studio~\cite{Curzi2024}, \textsc{sparta}~\cite{sion2018sparta}, and \textsc{cairis}~\cite{faily2018designing}.

Recently, this collection of available resources has been expanded to include more AI-specific resources, such as MITRE's \textsc{atlas}~\cite{ATLAS}, which complements \textsc{att\&ck} and focuses on AI-specific attack steps, and the OWASP 10 for LLMs~\cite{OWASPTTLLM}.
In addition to these resources, there are also more specific methodologies for eliciting threats in AI systems, such as attack trees for AI by Hoseini et al.~\cite{Hoseini2024}, the cloud security alliance's \textsc{maestro}~\cite{csamaestro,Huang2025} for threat modeling agentic AI, and frameworks for prioritizing AI vulnerabilities such as Microsoft's AI Bug Bar~\cite{AIBugBar}.
However, all these AI- or even GenAI-specific resources focus almost exclusively on the security aspects of these technologies.

\paragraph{Privacy threat modeling.}
With privacy threat modeling, the goal is to uncover privacy design flaws~\cite{Arvesen2022} or design weaknesses~\cite{De2016}.
In this area, the set of available frameworks, resources, and tools is much more limited.
One widely cited resource is Solove’s taxonomy of privacy~\cite{Solove2005}, which offers a structured classification of privacy harms and violations. 
It provides a conceptual foundation for reasoning about privacy and has been used to connect abstract threats to concrete harms in order to support prioritization~\cite{Cronk2021}. 
Another commonly referenced resource is the OWASP Top 10 Privacy Risks~\cite{OWASPTTPrivacy}, which highlights prevalent privacy vulnerabilities in software systems. 
However, while both resources identify and categorize privacy concerns, they do not provide a systematic methodology for analyzing a specific system architecture to uncover and model privacy threats.

\textsc{Priam}~\cite{De2016} is a privacy risk assessment methodology from De and Le Metayer.
This approach involves constructing harm trees, analogous to attack trees for security, with harm at the root, feared events in the tree, and weaknesses at the leaf nodes.
Although \textsc{priam} includes an information-gathering phase in which a system model is described, the methodology does not systematically analyze this model to derive privacy threats. 
Instead, the system representation serves as contextual input that supports the analyst's manual risk assessment.
\textsc{Mitre}'s \textsc{panoptic}~\cite{Shapiro2023} privacy threat model provides two taxonomies (contextual domains and privacy activities).
Finally, UsersFirst~\cite{Wang2024} is a more recent threat modeling framework that focuses specifically on user interactions to address threats such as unawareness, lack of control, and the identification of dark patterns in user interface design.
The only threat modeling resource that explicitly addresses privacy and AI is \textsc{plot4ai}~\cite{plot4ai}, a card deck of 138 AI-related threat cards.
It is broader in its focus, tackling all types of AI and also a broader range of threat categories, including (but not limited to) data \& data governance, cybersecurity, and safety \& environmental impact.
These taxonomies or knowledge resources, however, do not provide a methodology for eliciting concrete threats based on the system's design.

Among privacy threat modeling approaches that provide structured methodological support, \linddun{} remains one of the most established frameworks~\cite{Sion2025,Wuyts2015,deng2011privacy}. 
The \linddun{} acronym refers to seven privacy threat types:~Linking, Identifying, Non-repudiation, Detecting, Data Disclosure, Unawareness and Unintervenability, and Non-compliance. 
In \linddun{}'s threat knowledge base (often visualized as threat trees), these higher-level threat types are hierarchically refined into more specific \emph{threat characteristics}. These are contributing elements and conditions to a specific privacy threat. Furthermore, the knowledge base includes numerous concrete privacy threat examples which provide help to threat modelers in understanding the more conceptual elements of the framework.
Throughout its development~\cite{Wuyts2018}, and more explicitly in the most recent updates of the framework~\cite{Sion2025}, the ability to extend the framework has been an explicit objective.
\linddun{} is an open framework, allowing anyone to extend, fork or improve. 
It offers a threat knowledge base in the form of threat trees, and defines several methodological variants to guide systematic and structured privacy threat analysis. 

While \linddun{} enables structured privacy analysis for traditional software systems, it was not developed with the architectural characteristics of GenAI systems in mind. 
Consequently, it does not fully account for the system-level and model-specific privacy risks introduced by GenAI applications, as also highlighted by Savaliya et al.~\cite{Savaliya2026}.

\subsection{GenAI systems in practice}\label{ssec:backgenai}
A GenAI system is a software system that embeds generative models within a larger application architecture, including prompts, data flows, memory components, external tools, and user interfaces.
In contrast to discriminative machine learning models, which focus on predicting a target variable based on given features, a generative model is designed to generate data. 
Generative models are typically trained to learn a data-generating distribution that approximates the underlying structure of the training data.
During inference, the model is queried with a prompt or input, which activates latent representations to generate outputs based on conditional probabilities or learned data distributions.
GenAI models can be categorized by modality, including image~\cite{rombach2022high}, text~\cite{brown2020language}, audio~\cite{borsos2023audiolm}, video~\cite{openai_sora}, and tabular data~\cite{xu2019modeling}.

Several existing SoKs~\cite{shanmugarasa2025sok, hong2025sok, ma2025sok, smith2023identifying} cover privacy risks in GenAI.
Smith et al.~\cite{smith2023identifying} provide an early systematization of LLM attacks. 
They identify key dimensions that distinguish privacy attacks in LLMs and construct a taxonomy based on them. 
They categorize existing attacks and analyze emerging patterns. 
The work also surveys current mitigation strategies and discusses their pros and cons.

Ma et al.~\cite{ma2025sok} propose a lifecycle-oriented framework for analyzing semantic privacy risks in LLMs, focusing on how these risks arise during input processing, pretraining, fine-tuning, and alignment. 
The authors identify major attack vectors and examine how current defenses, including differential privacy, encrypted embeddings, edge-based computation, and unlearning, address these risks. Their analysis highlights significant gaps in semantic-level protection, particularly in contextual inference and the leakage of latent representations. 
They also outline several open challenges, including measuring semantic leakage, protecting multimodal inputs, balancing de-identification with generation quality, and improving the transparency of privacy enforcement.

Differing from Smith et al.~\cite{smith2023identifying}, and Nat et al.~\cite{ma2025sok}, which focus on the privacy risks of the training and fine-tuning data,
Hong et al.~\cite{hong2025sok} focus on LLM prompt security and offer several contributions. 
They introduce a multi-level taxonomy that organizes attacks, defenses, and underlying vulnerabilities. 
They formalize attacker models and cost assumptions that promote reproducible evaluation. 
They also release an open-source evaluation toolkit, \textsc{jailbreakdb}, which is an annotated dataset of jailbreak and benign prompts, and a comprehensive evaluation platform and leaderboard for state-of-the-art approaches.

Similar to~\cite{hong2025sok}, \cite{shanmugarasa2025sok} takes a broader look beyond the privacy risks of the training and fine-tuning data.  
\cite{shanmugarasa2025sok} mentions that previous surveys often give less attention to privacy threats arising from user interactions and advanced model abilities. 
The authors provide a broad analysis of privacy risks in LLMs and categorize them into four areas: (i)~privacy issues in training data, (ii)~privacy issues in user prompts, (iii)~vulnerabilities reflected in LLM-generated outputs, and (iv)~emerging privacy concerns involving LLM agents.

Except for \cite{shanmugarasa2025sok}, which adopts a system-level perspective on privacy within GenAI-integrated ecosystems, the other SoKs \cite{hong2025sok, ma2025sok, smith2023identifying} do not examine privacy risks from a holistic system perspective. 
Furthermore, none of these works explicitly identify the condition, dataflow, responsible party, the risk party, and the benefiting party in GenAI-related privacy attacks. 
This omission limits their applicability to practical privacy analysis of GenAI-based systems, where such distinctions are essential for industry practitioners. 

In this paper, we present a systematic categorization of the state-of-the-art GenAI privacy threats. We then structure and map these threats using the \linddun{} privacy threat knowledge base. 
The goal of this effort is \textbf{to make this specialized privacy threat knowledge and privacy engineering best practices accessible to software engineers developing GenAI-based systems}. \section{Methodology}\label{sec:methodology}

\begin{figure}
\includegraphics[width=\linewidth]{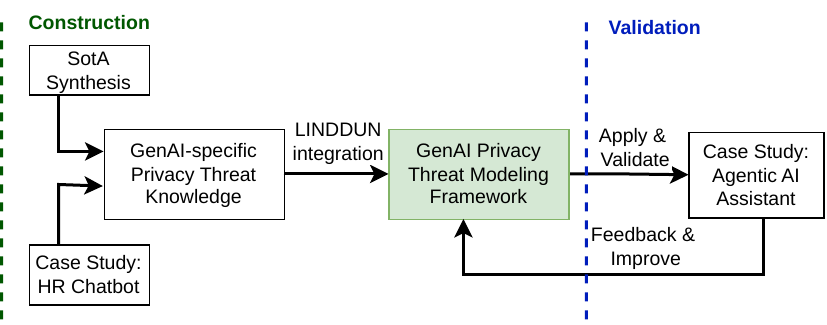}
    \caption{\label{fig:methodology}Overview of the approach.}
\end{figure}

To address RQ1 and RQ2, we define a novel privacy threat modeling framework that provides practical support to software engineers with identifying relevant GenAI-related privacy threats.
Our overall methodology consists of three steps. 

After selecting the foundation threat modeling framework (\cref{ssec:fwselection}), the following activities are performed:
(i)~\textbf{framework construction:}~the identification of the emerging, domain-specific privacy threats that are not adequately captured by generic privacy threat modeling frameworks when applied to GenAI–based applications (RQ1) and the integration of these in a new GenAI-specific privacy threat modeling framework (RQ2), and
(ii)~\textbf{framework validation:}~the evaluation of the framework in a contemporary and representative use case.
\Cref{fig:methodology} visualizes these two main methodology stages, which are further explained in \cref{ssec:derivation,ssec:framework-eval}.

\subsection{Foundation framework selection}\label{ssec:fwselection}

The development of a new framework from scratch is both time-consuming and prone to fragmentation, and often leads to threat models that are either too abstract to support practical reuse or too domain-specific to generalize beyond a single application context~\cite{sion2024leveraging}. 
Prior efforts~\cite{ouaissa2025framework, gholami2016advanced, gangavarapu2020target, alalade2025ptmf} that define entirely new threat modeling frameworks struggle with reuse, evolution, and integration into new workflows.

We select \linddun{} as the foundation for our framework because (i)~it is widely recognized as a mature framework and is actively used in both academia and industry, (ii)~it combines structured methodology with an extensible knowledge base, and (iii)~it is open for extension and refinement. 
Additionally, \linddun{} enjoys tool support by prominent threat modeling automation tools, such as OWASP ThreatDragon~\cite{OWASP2025ThreatDragon} and \textsc{sparta}~\cite{sion2018sparta}, which our new framework would inherit.

Instead of replacing existing infrastructure, we deliberately set out to strengthen and specialize it, enabling privacy-by-design practices to scale to complex GenAI systems without imposing unnecessary methodological overhead. Our approach promotes continuity, comparability, and reusability of threat knowledge while avoiding the reinvention of core threat modeling principles.

\subsection{Threat knowledge generation}\label{ssec:derivation}

The first stage of GenAI-specific privacy threat knowledge generation uses a combined bottom-up and top-down approach.
Part one (top-down) entails an extensive literature review on the SotA on LLM privacy threats and mapping these findings to the relevant \linddun{} privacy threat characteristics.
The second part (bottom-up) involves constructing an architectural representation of a GenAI-based chatbot application and subsequently analyzing the system's \linddun{} privacy threats.
Both of these parts serve as inputs to the creation of the GenAI Privacy Threat Modeling Framework.

\subsubsection{Top-down: systematic literature survey}
We performed a literature survey and identified 65 papers. A detailed overview of our search and selection protocol is provided in Appendix~\ref{app:search-protocol}.

\paragraph{Privacy knowledge generation.} 

Our synthesis begins with an in-depth review of selected papers to establish a foundational understanding to identify recurring GenAI usage paradigms. 
For each paradigm, we analyze reported privacy vulnerabilities and characterize them using \linddun{}. 
Novel paradigms or previously unreported vulnerabilities are flagged as candidates for extending the \linddun{} taxonomy. 
Each vulnerability is mapped to the relevant \linddun{} threat categories, along with its scenario and attacker model. 
This iterative workflow enables comprehensive coverage of existing research while supporting the identification of GenAI-specific privacy threats and potential updates to \linddun{}.Detailed results are provided in the supplementary materials.

\subsubsection{Bottom-up: HR Chatbot case study}\label{ssec:chatbot}
To generate GenAI-specific privacy threat knowledge in a bottom-up manner and identify gaps in the existing \linddun{} threat knowledge, we define a representative case study: a widely deployed GenAI-based HR chatbot. In this study, we use an open-source implementation~\cite{Bonifacio2023,Bonifacio2025} as the baseline, but many companies are developing similar HR chatbots~\cite{bamboohr,leenai}.

\paragraph{Privacy knowledge generation.}
We develop a detailed case description and construct a Data Flow Diagram (DFD) that captures its key components, data stores, and interactions. 
The DFD provides a holistic view of information flows within the chatbot ecosystem and serves as the basis for systematically identifying privacy threats.
Privacy threats are systematically elicited for the HR chatbot application by three threat modeling researchers using the traditional \linddun{} \textsc{pro} methodology~\cite{LINDDUNwebsite}, documenting each validated threat. The resulting threat list is then reviewed and validated in collaboration with two industry practitioners.The detailed case description, DFD, and threat list are provided as supplementary material~\cite{supplementaryMaterials}.

\subsubsection{Consolidation}
During threat elicitation for both the chatbot case study and the SotA review, we identified several previously unreported privacy threats. 
To ensure rigor, we held expert meetings with three threat modeling experts to examine each threat and reach consensus on its validity, relevance, and categorization. 
Particular attention was given to whether new threats required extending or refining the domain-specific \linddun{} framework. 
Inclusion decisions were made collaboratively. The finalized threat knowledge base was then reviewed and finalized in collaboration with two industry experts, to ensure an adequate reflection of the reality in practice.

\subsection{Evaluation of the framework}\label{ssec:framework-eval}
In the second stage, the GenAI Privacy Threat Modeling Framework is applied to an additional, more complex multi-agent GenAI application.
This way, the findings from the first stage are evaluated in a different application context to verify that they can be used across GenAI applications, and they are validated by three threat modeling experts from academia and two privacy experts from industry.
This process provides an empirical evaluation of the taxonomy's coverage and consistency.
\Cref{sec:validation} describes the outcome of this validation. \section{Framework construction}\label{sec:threats}

\Cref{ssec:sota-synthesis}  provides an overview of the privacy threats identified in literature analysis and the main findings from the chatbot case studies in~\cref{ssec:case}.
We address RQ1, compare the threats identified in the literature with those observed in practice, and discuss their implications in~\cref{ssec:observed-trends}.

\subsection{Top-down: SotA synthesis}\label{ssec:sota-synthesis}
Based on our synthesis of the literature, we describe four GenAI-based system paradigms, identify common attacker models (CAMs), and apply \linddun{} to these CAMs.

\subsubsection{GenAI-based system paradigms}
\label{sssec:sota-paradigms}
Users can interact with a GenAI-based system in four primary ways: 
(i)~direct use of a pre-trained (PT) model, 
(ii)~interaction with a specialized fine-tuned (FT) model, 
(iii)~use of a GenAI-based application, 
and (iv)~engagement with AI agent systems integrated with external tools. 
\Cref{fig:blackbox} illustrates these interaction modes and makes explicit the underlying data flows and control surfaces. 

The diagram can also be interpreted vertically as a lifecycle representation, although these phases are not strictly sequential. 
In practice, systems may skip stages entirely; for instance, a pre-trained model may be deployed directly without additional fine-tuning or application-layer integration.

In the first paradigm, the user interfaces directly with a PT foundation model (for example, a general-purpose chatbot such as ChatGPT) whose behavior is shaped by its large PT dataset and ongoing user interactions. 
User logs may be collected for model improvement if users have not opted out~\cite{openai_retention_policies}. 

In the second paradigm, the user interacts with an FT GenAI adapted to a specific domain or task (for example, a specialized HR or medical chatbot). 
Here, the model's behavior is derived from its FT dataset, and user logs continue to be collected in real time during interactions.
In the third paradigm, the user interacts with an LLM-powered application tuned with system prompts (for example, templates used to generate medical notes).

In the fourth paradigm, the user interacts with a GenAI-based application that is embedded within a broader software ecosystem, such as an operating system assistant. 
In this setting, the model may invoke third-party tools or Retrieval-Augmented Generation (RAG) components as part of an agent-based configuration. 
More advanced deployments further adopt multi-agent architectures in which multiple GenAI-based agents interact. 
These agentic designs introduce an expanded threat surface, as sensitive information can propagate across components and between interacting models.

\subsubsection{Common Attacker Models in literature}
\label{sec:common-tm}
\begin{figure}
    \includegraphics[width=\linewidth]{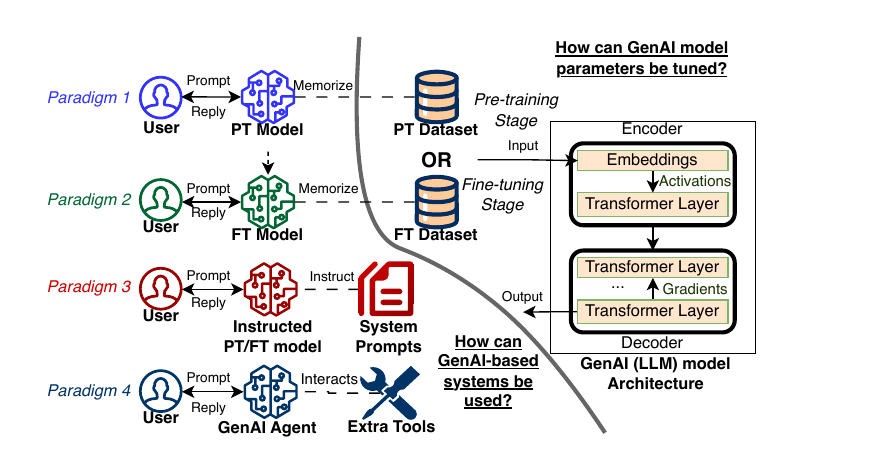}
    \caption{\label{fig:blackbox}GenAI-integrated systems.}

\end{figure}

\begin{table*}[]
{
\begin{threeparttable}
\caption{\label{tab:tm_from_sota}Overview of Common Attacker Models (CAMs) identified in the literature and corresponding citations. }
\renewcommand{\arraystretch}{0.95}
\newlength{\colw}
\setlength{\colw}{3cm}
\small
\centering
\begin{tabular}
{l M{1.5cm} R{1.6cm} l p{3cm} p{8.5cm} }
\toprule
~&\multicolumn{1}{l}{\textbf{CAM}}              & \textbf{Found in}
& \textbf{Actor} 

& \textbf{Conditions} 
& \textbf{Impact}    \\
\midrule
\multicolumn{5}{l}{\textbf{User-GenAI System Threats}}\\
\cmidrule{2-6}
& CAM1 
& \pctbar{13} 
& System
& Access to system input
& \UI ~\cite{wang2024dye4ai,yu2024privacy, hong2023dp, duan2023flocks, weiss2024your, staab2023beyond}
\DA ~\cite{zhao2025llms, wen2024membership}
\\
& CAM2 
& \pctbar{45}
& User
& Access to system output
& 
\SP~\cite{hui2024pleak}
\PTD~\cite{wang2023decodingtrust, fu2024membership,  maini2024llm, lukas2023analyzing, liu2024autodan, carlini2022quantifying, sadasivan2024fast, pinto2024extracting, carlini2021extracting, wei2023jailbroken,  kim2023propile, kang2025can, wang2024property, he2025towards, huang2022large, santurkar2023whose,xie2024differentially, shen2024anything, yu2023bag} 
\FTD~\cite{jayaraman2024combing, deng2024masterkey, li2025differentiation, akkus2025generated, xin2025false, li2025odysseus}
\\
\midrule
\multicolumn{5}{l}{\textbf{Cross-Boundary Threats}}\\
\cmidrule{2-6}
& CAM3 
& \pctbar{3} 
& FT Party
& Access to system output
& \PTD ~\cite{chen2024janus, li2024shake}
\\
& CAM4 
& \pctbar{4}
& System
& Agent permissions 
& \FTD ~\cite{liu2024precurious}
\OD ~\cite{debenedetti2024privacy}
\\
& CAM5 
& \pctbar{13}
& Agent
& System permissions 
& \UD \cite{shi2025prompt, wang2025obliinjection, 10.1145/3719027.3744840}
\OD ~\cite{shao2024privacylens,liu2024demystifying, bagdasarian2024airgapagent, cui2025odyssey, greshake2023not}
\\
\midrule
\multicolumn{5}{l}{\textbf{Within-system Threats}}\\
\cmidrule{2-6}
& CAM6 
& \pctbar{24}
& Client
& Access to intermediate computations
& \PTD~\cite{10.1145/3719027.3765174, 10.1145/3719027.3744820, feng2024uncovering, ferrando2024know, reizinger2024cross, tan2026wasmydata,song2020information, pan2020privacy, patil2024can}
\FTD~\cite{chen2024unveiling, mai2023split, zhang2023dpzero}
\UI~\cite{luo2025shadow, wu2025know}
\\
\bottomrule
\end{tabular}
\end{threeparttable}
\begin{tablenotes}[online]
\scriptsize
\vspace{-0.3cm}
\begin{multicols}{2}
\item[\UI] are textual prompts, images, and speech inputs that may embed private user information.
\item[\SP] may embed private or proprietary information.
\item[\DA] such as personal preferences, are inferred through (subsequent) user-GenAI system interactions.
\item[\PTD (\FTD)] refers to information belonging to the pre-training (fine-tuning) datasets.
\item[\OD] are information obtained from external applications, databases, agents, or tools.
\item[\UD] are user documents, user logs and actions.
\end{multicols}
\end{tablenotes}
}
\end{table*}

Through an extensive literature survey and analysis, we identify six main types of leakage that can be exploited across different paradigms or stages of a GenAI-based system, which we term Common Attacker Models (CAMs) (shown in \cref{tab:tm_from_sota}).
We differentiate and categorize attacker models based on two criteria:  
(i)~the type of targeted information (i.e., system inputs, system outputs, or files and data accessible through system permissions),
and (ii)~the type of actor conducting the attack or exploiting the leakage, including individuals (e.g., malicious users) and organizations (e.g., fine-tuning or pre-training parties). 
Hence, the CAMs are defined and named as: 
(i)~user-to-system leakage; 
(ii)~system-to-user leakage; 
(iii)~PT-to-FT leakage; 
(iv)~system-to-agent leakage; 
(v)~agent-to-system leakage; 
and (vi)~residual privacy leakage.

Following the severity prioritization proposed by Kunz et al.~\cite{kunz2023privacy}, we categorize the threats emerging from each CAM into three groups: (i)~user-to-system data flows, (ii)~cross-boundary data flows, and (iii)~within-system threats, including insider risks.
CAM1 and CAM2 involve threats that emerge from direct \textbf{user-system interactions}. 
CAM3, CAM4, and CAM5 involve \textbf{cross-boundary threats}: in CAM3 and CAM4, the fine-tuning and pre-training systems belong to different, potentially conflicting parties that may act maliciously toward one another, while in CAM5, privacy threats arise when an AI agent interacts with third-party tools or other AI agents. 
Finally, CAM6 pertains to \textbf{within-system threats} because intermediate computations are typically transmitted in federated learning scenarios, where all parties operate within a semi-trusted environment.

\paragraph{CAM1: User-to-System Leakage.}
A (malicious) GenAI system may collect sensitive information or elicit additional details from user inputs (user queries, uploaded documents, uploaded photos, etc.) by 
(i)~strategically manipulating its responses, for example, by asking follow-up questions such as ``Can you be more specific?'' 
or (ii)~by presenting itself as a trustworthy entity. 
Such behavior may arise if the underlying system is 
(i)~privacy-indifferent, honest-but-curious, 
or (ii)~intentionally malicious, for instance, when system prompts are deliberately designed to solicit more information than necessary, 
or (iii)~compromised by security attacks, such as through the insertion of a backdoor.
In this attack model, the victims are users or data subjects whose information appears in the application context through user prompts or contextual data during their interactions with the GenAI system.
A growing body of work supports that LLMs can infer sensitive personal attributes~\cite{wang2024dye4ai}, personal preference~\cite{zhao2025llms} and contents~\cite{yu2024privacy, hong2023dp, duan2023flocks, weiss2024your, staab2023beyond} from user prompts.

\paragraph{CAM2: System-to-User Leakage.}
A (malicious) user may receive GenAI outputs that inadvertently reveal 
i)~pre-training (PT) data,
ii)~fine-tuning (FT) data, 
iii)~previous user interaction logs, 
iv)~or system-level knowledge, such as system prompts, thereby leaking private information about data subjects. 
This privacy risk may be amplified through crafted attacks such as Membership Inference Attacks (MIA), Attribute Inference Attacks (AIA), or Data Reconstruction (DR).

In a MIA~\cite{fu2024membership}, the (malicious) user seeks to determine whether a specific individual's data was included in the training dataset, FT dataset, system prompts, or user interaction logs.
In AIA, the attacker uses model replies to infer additional sensitive attributes about a target individual beyond what is explicitly disclosed. 
Even if the system does not directly reveal protected data, correlations encoded in the model may allow attackers to infer hidden attributes from seemingly benign responses.
This can occur when carefully crafted prompts cause the model to generate replies that reflect memorized patterns, reveal fragments of prior in-context inputs~\cite{wen2024membership}, or expose sensitive information embedded in system prompts (e.g., medical notes)~\cite{hui2024pleak}. 
Such attacks may leverage crafted queries~\cite{kim2023propile}, jailbreak techniques that bypass guardrails~\cite{wei2023jailbroken}, or automated large-scale prompt generation~\cite{liu2024autodan, deng2024masterkey}.

DR goes further by recovering portions of sensitive records from model outputs. 
Attackers may analyze multiple responses, confidence scores, or loss signals~\cite{fu2024membership} to iteratively reconstruct training data. 
Techniques such as generation-and-ranking strategies (``bag of tricks'')~\cite{yu2023bag} can improve reconstruction quality. 
Moreover, cross-modal and cross-lingual leakage has been demonstrated~\cite{pinto2024extracting}, where outputs that appear harmless in one modality (e.g., text) enable reconstruction of sensitive information from another modality (e.g., images), effectively exposing missing or redacted fields.

\paragraph{CAM3: PT-to-FT Leakage.}
A (malicious) fine-tuning party may deliberately orchestrate the fine-tuning process to reconstruct data records from the pre-training dataset. 
This attack assumes that some pre-training records are partially forgotten due to catastrophic forgetting~\cite{chen2024janus, li2024shake}. 
Carefully designed fine-tuning can help the malicious model recall partially forgotten information and reinforce retention of previously learned data.

\paragraph{CAM4: System-to-Agent Leakage.}
Integrating an insecure or unauthorized GenAI model as a component into a local application can cause several privacy issues.
A malicious pre-training party that implants a backdoor can facilitate the reconstruction of records from the fine-tuning dataset once the fine-tuning party uses the backdoored model as a foundation model~\cite{liu2024precurious}. 
Similarly, if a compromised GenAI model is deployed within an application and granted access to user files and application data, it may enable the extraction of sensitive information from the application operation context~\cite{ debenedetti2024privacy}.

\paragraph{CAM5: Agent-to-System Leakage.}
A GenAI agent may voluntarily, due to a malfunction or poor design, or be misled into performing actions that expose users' sensitive files, interaction logs, and/or conversation histories to 
(i)~external systems, 
(ii)~external tools, 
(iii)~other AI agents, 
or (iv)~security attackers who have gained access by compromising the GenAI-based system.

For example, indirect prompt injection allows adversaries to embed malicious instructions into retrievable content, leading GenAI-integrated applications to exfiltrate user data such as credentials, personal information, or chat histories~\cite{greshake2023not, wang2025obliinjection}. 
Similarly, AI agents with tool-use capabilities may access personal resources (e.g., medical conditions, calendars, credit card numbers, etc.) and transmit or infer sensitive information without the user's awareness~\cite{shao2024privacylens, cui2025odyssey}.
Moreover, membership inference attacks against RAG systems can determine whether specific documents are present in the datastore using carefully crafted natural-language queries~\cite{10.1145/3719027.3744840}.

\paragraph{CAM6: Residual Privacy Leakage.}
When system memories or the computations of training and inference are shared among multiple participating clients, those clients may obtain white-box access or access to the model’s intermediate computations, which can expose residual privacy risks.
A typical white-box scenario is illustrated in the right-hand side of~\cref{fig:blackbox}.

Intermediate computations include various types of internal model signals that may leak sensitive information. 
First, embeddings are vector representations of tokens, and prior work has shown that they can be inverted to recover the original text~\cite{song2020information} or even the original prompts in Mixture-of-Experts (MoE) architectures~\cite{10.1145/3719027.3765174}, thereby enabling reconstruction attacks~\cite{song2020information, 10.1145/3719027.3765174}.
Second, forward activation values capture layer-by-layer transformations within the model, and these transformations can be inverted to reveal properties of the input~\cite{chen2024unveiling, 10.1145/3719027.3744820} in Collaborative Learning settings. 
Third, gradients and losses produced during backpropagation may leak information about individual records, as demonstrated in recent studies on gradient leakage~\cite{feng2024uncovering}. 
Third, model weights encode the model's learned representations and may also reveal characteristics of the underlying data~\cite{patil2024can}.
Finally, in a multi-tenant LLM serving framework, the user Key-Value (KV) cache can lead to the re-identification of user input~\cite{luo2025shadow, wu2025know}.

\subsubsection{Applying LINDDUN to CAMs}
\label{sec:mapping}

\begin{table*}
{\footnotesize
\centering
\newlength{\colwidth}
\setlength{\tabcolsep}{4pt}
\renewcommand{\arraystretch}{0.95}
\setlength{\colwidth}{2.62cm}
\small
\caption{\label{tab:linddun_vul}\linddun{} analysis of the 6 common attacker models (CAMs) in GenAI-based systems.}
\begin{threeparttable}
\begin{tabular}{l 
p{\colwidth} 
p{\colwidth} 
p{\colwidth} 
p{\colwidth} 
p{\colwidth} 
p{\colwidth}}
\toprule 
& \raggedright\textbf{\footnotesize CAM1: User-to-\\System Leakage}
& \raggedright\textbf{\footnotesize CAM2: System-to-\\User Leakage}
& \raggedright\textbf{\footnotesize CAM3: PT-to-FT Leakage}
& \raggedright\textbf{\footnotesize CAM4: System-to-\\ Agent Leakage}
& \raggedright\textbf{\footnotesize CAM5: Agent-to- \\System Leakage}
& \raggedright\textbf{\footnotesize CAM6: Residual Privacy Leakage}\tabularnewline
\midrule  
\multicolumn{7}{l}{\textbf{Linking:} Associating data items or user actions to learn more about an individual or group.}\tabularnewline
\arrayrulecolor{gray}
\cmidrule{2-7}
\arrayrulecolor{black}

 & Inputs contain linkable attributes that allow associating information to the user or other data subjects.

 & System outputs to a specific user contain linkable information pertaining to other data subjects.

 & Outputs to the fine-tuning party include linkable information of the data subjects in the PT dataset. 

 & Outputs to agent include linkable information about data subjects from the application context.

 & Uploaded files reveal stylistic or contextual signals that allow linking multiple user sessions or sensitive information. 

 & Intermediate computations reveal patterns that link additional data to individuals or groups. 

\tabularnewline

\midrule 
\multicolumn{7}{l}{\textbf{Identifying:} Learning the identity of an individual, through leaks, deduction, or inference.}\tabularnewline
\arrayrulecolor{gray}
\cmidrule{2-7}
\arrayrulecolor{black}

 & The system receives explicit PII$^{1}$\ or QIDs$^{2}$ from inputs.

 & System outputs leak PII$^{1}$ or QIDs$^{2}$ from logs or metadata.

 & System outputs leak PII$^{1}$ or QIDs$^{2}$ from pre-training records.

 & System outputs leak PII$^{1}$ or QIDs$^{2}$ from application context.
 & Agent leaks PII$^{1}$ or QIDs$^{2}$ from accessed information.

 & Intermediate computations manifest identifying indicators.
\tabularnewline
\midrule 
\multicolumn{7}{l}{\textbf{Non repudiation:} Being able to attribute a claim to an individual.}\tabularnewline
\arrayrulecolor{gray}
\cmidrule{2-7}
\arrayrulecolor{black}
 & Stored logs prevent users from denying disclosure of sensitive data.

 & Stored logs prevent the system from denying disclosure of sensitive data.

 & Stored logs prevent the system from denying disclosure of sensitive data.

 & Stored logs prevent the system from denying disclosure of sensitive data.
 & Exposed file contents reveal actions or claims that agent cannot deny.

 & Intermediate computations disclose traces of user information or actions.\tabularnewline
\midrule 
\multicolumn{7}{l}{\textbf{Detecting:} Deducing the involvement of an individual through observation.}\tabularnewline
\arrayrulecolor{gray}
\cmidrule{2-7}
\arrayrulecolor{black}
 & 
 The system detects activities or user intent based on inputs.
 & 
User detect whether private information is included or excluded based on system output.
& 
 Service providers detect whether a record or a data subject's information was used in training.

 & Agent detects whether data subject's information was used in the application.

 & User actions can be detected from file access patterns or agent logs.

 & Observing intermediate computations allows inference of participation or record membership. \tabularnewline
\midrule 

\multicolumn{7}{l}{\textbf{Data disclosure:} Excessively collecting, storing, processing or sharing personal data.}\tabularnewline
\arrayrulecolor{gray}
\cmidrule{2-7}
\arrayrulecolor{black}
 & User inputs contain sensitive information that providers can store or repurpose.
 & 
 System outputs reveal stored logs, system prompts, or cross-modal information.
 & 
The fine-tuning party may extract targeted pre-training data using curated triggers.
 & 
 The agent may extract targeted application data through pre-planted triggers.

 & Agents may disclose raw file contents or summaries to external parties.

 & Intermediate computations may enable partial or full reconstruction of sensitive information.
\tabularnewline

\midrule 
\multicolumn{7}{l}{\textbf{Unawareness and Unintervenability:} Insufficiently informing, involving or empowering individuals in the processing of their personal data.}\tabularnewline
\arrayrulecolor{gray}
\cmidrule{2-7}
\arrayrulecolor{black}
  & Users lack control over system logging of their inputs and outputs.

 & Users lack knowledge of or control over system log retention, usage, or system outputs.

  & 
  Users cannot observe or prevent the inclusion of their data in training.

 & 
 Users cannot observe or prevent the exposure of their data to agent.

 & 
 Users do not understand how agents store, forward, or process their files.

 & 
 Data subjects are unaware their data persists in intermediate computations \& cannot access/correct.\tabularnewline
\midrule 
\multicolumn{7}{l}{\textbf{Non-compliance:} Deviating from security and data management best practices, standards and legislation.}\tabularnewline
\arrayrulecolor{gray}
\cmidrule{2-7}
\arrayrulecolor{black}

 & Illegitimate harvesting of sensitive query information violates consent and lawful basis requirements.

 & Data retention without consent or leakage of other users’ data violates regulations.

 & Unauthorized dataset extraction violates privacy regulations.

 & Unauthorized dataset extraction violates privacy regulations.

  & Forwarding data to external parties can violate confidentiality and purpose-limited rules.

 & Leakage through derived representations may conflict with data
minimization requirements.
\tabularnewline
\bottomrule 
\end{tabular}
\end{threeparttable}
\begin{tablenotes}[online]
    \small
    \item[$^1$PII:] personally identifiable information
    \item[$^2$QID:] quasi-identifier
    
\end{tablenotes}
}
\end{table*}

 Mapping the Common Attacker Models (CAMs) to LINDDUN provides two main benefits: (i)~it enables more systematic analysis and mitigation by revealing how a single CAM may span multiple LINDDUN threat categories;
and (ii)~it exposes limitations in LINDDUN, highlighting where GenAI-specific risks diverge from traditional software privacy threats. The full mapping is shown in~\cref{tab:linddun_vul}.

\subsection{Bottom-up: practical case study}
\label{ssec:case}
\paragraph{Case description.}~Information chatbots are a common application of LLMs, enabling interactive query–response interactions with users. In this study, we focus on a generative Human Resources (HR) chatbot designed for internal organizational use. The chatbot assists employees in accessing information about company policies and their employment data.
For example, a user may enter prompts such as ``\emph{How many vacation days do I have left?}'' or ``\emph{What is the company policy on unused vacation days?}''. It implements several core functionalities to achieve this, such as authentication, text/voice interaction, RAG, and internet search.

\paragraph{Threat elicitation.}~Our threat elicitation for the chatbot identified 127 privacy threats.  Below, we highlight examples of representative threats from each category. The detailed case description, DFD, and threat list are provided in the supplementary materials~\cite{supplementaryMaterials}. 
\begin{description}
   \item[User-GenAI threats] Users may overshare personal information in their queries due to uncertainty about what is required, thereby unnecessarily increasing the chatbot’s exposure to sensitive data.
   \item[Cross-Boundary threats] When user queries are forwarded to external AI service providers, excessive personal data may enable profiling across requests. Such inferred information could be repurposed for secondary uses, such as advertising or resale.
   \item[Within-System threats] System logs that retain voice inputs, transcriptions, and session identifiers may prevent users from plausibly denying sensitive requests, exposing them to reputational or professional risks.
\end{description}

\subsection{Answering RQ1: GenAI privacy threats}\label{ssec:observed-trends}
As shown in Table~\ref{tab:tm_from_sota}, most academic work analyzes privacy leakage in GenAI primarily from the perspective of user–GenAI interactions. 
In practice, however, GenAI systems operate within complex enterprise environments where multiple components interact, introducing additional privacy risks through cross-boundary data flows and system-component interactions.

Our findings indicate that many privacy risks in GenAI systems remain well captured by existing \linddun{} categories, particularly \emph{Linking}, \emph{Identifying}, \emph{Non-repudiation}, and \emph{Detecting}. 
These categories sufficiently explain several threats identified in both the literature and our case study. 
However, in CAM6, where intermediate computations themselves constitute the exposed data flows, substantial extensions to \linddun{} are required. 
Although such internal signals may not appear linkable or identifiable at first glance, advanced analysis can render them indirectly linkable.
In addition, the memorization properties of GenAI models amplify concerns related to \emph{Unawareness}, as personal data may persist far longer than users anticipate. 
Taken together, these observations expose a GenAI-specific threat surface that is not fully addressed by the current \linddun{} framework. 

We next discuss the key trends that emerged from our case studies and literature analysis.

\subsubsection{Stochasticity}
Many emerging GenAI-related privacy threats stem from the probabilistic nature of AI systems. 
Because the reply given to a certain prompt is always different and largely autonomously `decided' by the model itself, it is difficult to predict to output of the system. This unpredictability leads to a series of important potential harms, such as (i)~users (or even other AI-based agents) acting upon the erroneous output from an AI system, or 
(ii)~the AI system misinterpreting a prompt and context, and sharing (sensitive) data about one person with another.

\subsubsection{AI Literacy}
We note a significant number of new threats related to so-called `AI literacy'~\cite{Long2020}. 
Many end users of AI technologies do not possess the technical expertise to fully understand the mechanisms through which such systems function. 
This, combined with the fact that many such systems are interacted with as if having a conversation with another person, makes it easy for end users to start treating the AI system as if the system is another person. 
Users who see an AI-based system as `magically' answering their questions, treating it similarly to asking a friend for advice, may not be aware of the many pitfalls and privacy risks related to such a system. We see this most clearly in a substantial addition of new threat characteristics to the \emph{Unawareness and Unintervenability} threat type (see \cref{sec:threatknowledge}), but AI literacy also permeates many of the other newly added tree nodes and examples (e.g., the end user sharing too much of their sensitive personal data with the AI system under the \emph{Data Disclosure} threat type).

\subsubsection{Manipulation}
The conversational nature of GenAI systems, combined with a lack of AI Literacy among users, and a strong ability of GenAI systems to infer new information and match patterns lead to such systems potentially taking advantage of people by saying what the person wants to hear, echoing their own biases back to them. This is further exacerbated by many GenAI systems' unwillingness to say `no' or to go against what the user is saying, which can lead to manipulative behavior (either unintentionally, or intentionally). Some recent examples of such behaviors are GenAI chatbots which push their users towards taking drastic decisions (such as breaking up with a long-term partner~\cite{Fike2025}), or to take harmful actions (such as acting upon urges to self-harm, or even to commit suicide~\cite{Pichowicz2025,DeFreitas2023,Chatterjee2025,Kuenssberg2025}). In a more advanced scenario, a chatbot could also make certain claims and then deny having made them at a later point in the future. For example, in the case of the HR chatbot, the system could claim that a user's holiday was approved at one point in time, but deny ever having said this at another point in the future, effectively `gaslighting' the employee on whether a previous conversation ever really happened. We capture this new `branch' of threats in the newly added threat characteristic U.3 (see \cref{sec:threatknowledge}).

 \section{Framework validation}\label{sec:validation}
We validate the GenAI Privacy Threat Modeling Framework by applying it in a second case study. We first define and motivate the choice of the case, then present the outcome of our validation.

\subsection{Agentic AI Assistant case study}
\paragraph{Case description.} The multi-agent AI Assistant case represents a complex application scenario in which a series of autonomous AI agents perform tasks on behalf of the user across multiple systems. Operating on a terminal device such as a smartphone or laptop, a local agent can interact directly with graphical interfaces or invoke application-level actions through inter-process communication mechanisms. It is also connected to cloud-based agents capable of querying third-party APIs or external agent providers. Typical usage scenarios include answering user queries, booking travel or accommodation, ordering food, composing messages, and maintaining personalized memories. Because the agent has authorization to access private data and execute actions autonomously, it presents a substantially larger attack surface and a richer set of potential threats. The full case description is provided in the supplementary materials~\cite{supplementaryMaterials}.

This case was chosen for its intrinsic complexity, potential for high-impact threats, and most importantly, accurate reflection of popular GenAI-based applications that are being built today~\cite{Saleem2025}.
Furthermore, the case was validated with industry experts that are actively building such applications, which ensures both realism and relevancy. 

\paragraph{Threat elicitation.} 
The construction of the DFD and elicitation of privacy threats was done analogous to the HR chatbot case study (see \cref{ssec:chatbot}). 
First, the Data Flow Diagram (DFD) was modeled and then thoroughly analyzed by a team of three privacy threat modelers based on the new GenAI Privacy Threat Modeling Framework to elicit privacy threats. Then, the resulting list of threats was discussed with and reviewed by two industry practitioners. 
During this process, specific attention was paid to any gaps in the framework's threat knowledge. For efficiency reasons, this threat elicitation phase was scoped to exclude interactions that are not GenAI-related (e.g., telemetry) or that were already analyzed in the chatbot case study.

The resulting threat list consists of $98$ privacy threats, of which $9$ (or around $ \sim9\%$) are based on a newly introduced GenAI-specific threat characteristic.
The detailed threat list can be found in the supporting materials~\cite{supplementaryMaterials}.

\subsection{Validation outcome}

We have validated the case study through expert review by the industry practitioners, who are considered domain experts on the use and integration of Agentic AI in a real-world context.

We first note that no new threat characteristics needed to be added to the framework in this validation step, which suggests that the new threat knowledge is sufficiently general and applicable for privacy threat modeling of other GenAI-based applications. The feedback loop of this validation phase led to the inclusion of an additional 36 Agentic examples to the knowledge base.

Furthermore, we highlight the relatively low incidence of GenAI specific threat characteristics, which is only $9\%$. This confirms the importance of our decision to refine existing frameworks and reuse existing knowledge efforts, rather than to build a purely domain-specific framework from scratch. \section{GenAI privacy threat knowledge base}\label{sec:threatknowledge}
This section describes the resulting privacy threat knowledge base for GenAI. We first elaborate on the new threat characteristics and then discuss the new GenAI threat examples\footnote{The full threat trees of the GenAI privacy threat modeling framework, including all new threat characteristics and examples, are available as supplementary material~\cite{supplementaryMaterials}.}.

\begin{figure}
    \includegraphics[width=.975\linewidth]{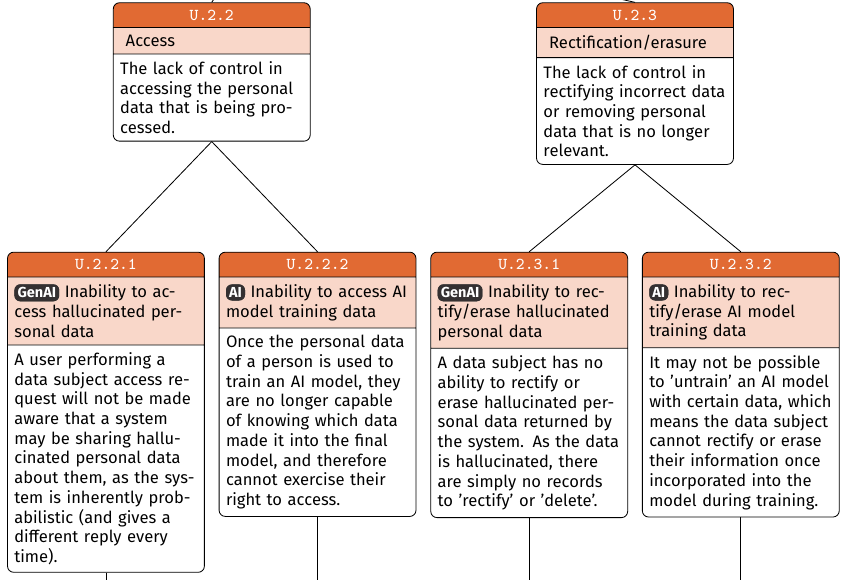}
    \multilinecaption{\label{fig:access-rectification}}{Example unawareness \& unintervenability tree.}{The `U.2.2 Access' and `U.2.3 Rectification/erasure' threat characteristics were split and refined into AI- and GenAI-specific sub-characteristics.}
\end{figure}

\subsection{Threat characteristics}
The most significant changes to the original \linddun{} privacy threat knowledge base are situated in the Data Disclosure and Unawareness and Unintervenability threat types.
\begin{description}[wide]
\item[{Data disclosure.}] The new threat knowledge adds two new threat characteristics to the Data Disclosure threat type. A first new threat characteristic is \emph{DD.1.3 Data Type Structure}. It reflects the privacy risks tied to the reversibility of certain intermediary data structures often used within GenAI systems. The literature has shown that such intermediate computations can sometimes be leveraged to reveal sensitive personal data that was part of a fine tuning or training data set. The second new threat characteristic, \emph{DD.3.5 Fabrication}, captures threats related to the tendency of GenAI systems to hallucinate information. This poses important privacy consequences, as false personal data could be treated as if it is accurate.

\item[{Unawareness \& unintervenability.}] This threat type was extended with a total of five new characteristics. The existing \emph{U.2.2 Access} and \emph{U.2.3 Rectification/erasure} threat characteristics were each split and refined into two new, AI-specific characteristics to reflect the specific challenges to accessing, rectifying, or erasing training data or hallucinated personal data from such systems. An entirely new characteristic is \emph{U.3 Interference with personal decision making}, which captures the new category of privacy threats as described under `Manipulation' in \cref{ssec:observed-trends}.

\item[{Non-compliance.}]
Finally, there are two additional characteristics for the Non-compliance threat type, one focusing on regulatory non-compliance with the EU AI act (Nc.1.3) and another on non-compliance with AI standards and best practices (Nc.4.2).

\end{description}
\subsection{Examples}
The analysis of the two application cases led to an extensive collection of \casethreatcount{} privacy threats.
These threats are distilled into concrete examples for all the characteristics in the \linddun{} threat trees in order to assist threat modelers in translating non-specific characteristics to their GenAI applications.
Where relevant, the examples were generalized to be applicable to the entire GenAI domain, or even to AI in general.
The resulting GenAI privacy threat knowledge base contains \excount{} examples (with AI, GenAI, Chatbot, or Agentic domain tags).
All characteristics in the  \linddun{} threat knowledge base (even those that are not exclusive to GenAI) were extended with examples, with the exception of the Non-compliance tree and one Non-repudiation threat characteristic (Nr.1.2).

\subsection{Answering RQ2: Practical considerations}
We have opted to build the GenAI privacy threat modeling framework on top of the already well-established \linddun{} knowledge base in order to minimize friction for adoption in practice. In addition, several implementation choices were made to increase the usability of the framework, which we highlight below.

\begin{figure}
\begin{minipage}{.63\linewidth}
\includegraphics[width=1\linewidth]{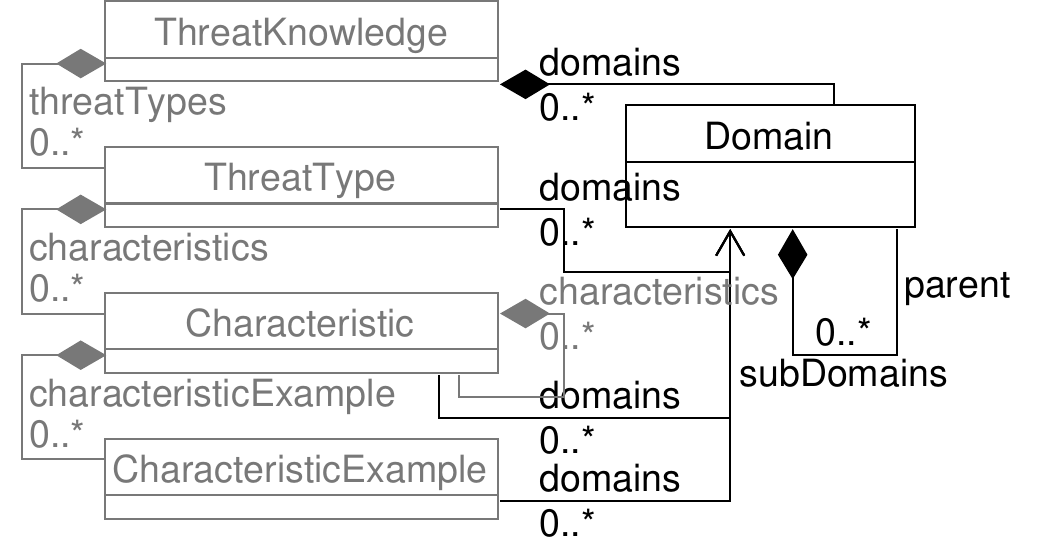}
\end{minipage}
\begin{minipage}{.33\linewidth}
    \begin{forest}
[General
    [\ldots[,phantom]]
    [AI [GenAI
            [Chatbot]
            [Agentic]]
        [ML]]]
\end{forest}
\end{minipage}
\multilinecaption{\label{fig:domain-structure}}{Domain metamodel and hierarchy.}{This diagram illustrates how the \linddun{} threat knowledge metamodel (gray) is extended with a domain-class that can be referred to across the threat knowledge tree. The right-hand side shows how GenAI is encoded in the domain hierarchy.}
\end{figure}

\paragraph{Domain hierarchies.}~The \linddun{} privacy knowledge base and code to generate the different formats (e.g., threat trees, documentation, \linddun{} \textsc{go} cards) are made available on the project website~\cite{linddunTechStack,Sion2025}. In order to build our domain-specific framework, we have extended \linddun{}'s meta-model~\cite{Sion2025} to support hierarchical domain tags for threat characteristics and examples. This enables the encoding of domains and subdomains into the threat knowledge, not just for (Gen)AI in our use case, but also other (sub)domains. Concretely, this opens the door for other domain-specific variants of \linddun{} to be created (e.g., for specific sectors, such as finance or eHealth). \Cref{fig:domain-structure} shows the extension to the meta-model as well as the used domain hierarchy for the GenAI privacy threat modeling framework.

\paragraph{Filtering.}~To provide threat modelers with the information from the privacy threat knowledge base which is relevant to them, we have adapted the code used for the generation of the threat trees to support filtering based on the newly introduced domain tags. This enables the generation of domain-specific threat trees which only include domains relevant to the desired use case at the push of a button.

\paragraph{Future-proofing.}~The field of software engineering rapidly and continuously evolves. Therefore, it is critical that the privacy threat knowledge base is kept up-to-date along with it, as new privacy threats emerge and existing threats become better understood. By making the GenAI-specific privacy threat modeling framework publicly available~\cite{supplementaryMaterials}, along with the code necessary to generate the different knowledge representations, we aim to future-proof the framework for continuous updates by academia and practitioners.

 \section{Threats to validity}\label{sec:ttv}
This section discusses threats to internal and external validity.

\subsection{Internal validity}

\paragraph{Model construction.}
The model construction was done by privacy engineering researchers. This means the models may not be representative for practitioners that would construct these models. This was partially addressed through the involvement of two privacy experts from within the industry to address any ambiguities in the description of the models.

\paragraph{Threat analysis.} 
The analysis of the shortcomings in the existing generic \linddun{} privacy threat knowledge is performed through the execution of interaction-based \linddun{} privacy threat elicitation (\linddun{} \textsc{pro}).
This is only one method of performing privacy threat elicitation; other approaches~\cite{Wuyts2020,VanLanduyt2025} may have an impact on the shortcomings that were identified in the privacy threat knowledge.
However, given that the interaction-based threat elicitation approach is one of the most fine-grained ones, it does allow for the detailed application of the \linddun{} threat characteristics.

To further address the limitation of relying only on \linddun{} threat knowledge, our approach included the analysis of the state of the art on security and privacy threats in GenAI to further complement and extend the findings.

\subsection{External validity}

\paragraph{Case selection.} The construction of GenAI-specific privacy threat knowledge depends on the selection of a relevant and representative application cases to analyze. In collaboration with industry practitioners, we selected two cases: a chatbot and an agentic assistant. These application cases may not cover all potential scenarios where application developers may integrate GenAI components into their applications. However, given the prominence of chatbot-like interfaces for users to interact with these systems and the agentic assistant for task delegation, the selected case studies do cover two very prominent and broad categories of GenAI usage.

\paragraph{Threat knowledge generalization.}
A final threat to validity is the generalization of the threat knowledge for other types of GenAI-based application which more fundamentally differ from the two case studies that were used to construct and evaluate our new framework.
This can be addressed in the future by continuing to apply this threat knowledge on other applications as well.
As part of the encoding of the domain-specific threat knowledge in the \linddun{} data structure, the characteristics and examples were generalized as much as possible to ensure broader applicability. We observe that 40\% of the concrete threat examples that were included in the threat knowledge can be generalized to be applicable beyond the specific chatbot or agentic domain. We do note that, given the fast-paced evolution in this domain, threat knowledge on GenAI will continue to be a moving target.
The existing \linddun{} threat knowledge and tooling~\cite{linddunTechStack,Sion2025} together with our meta-model extension do enable the continuous update and evolution of this threat knowledge.

 \section{Conclusion}\label{sec:conclusion}

The growing adoption of GenAI across applications brings significant privacy risks. 
Although considerable research exists on security flaws and tools for analyzing security threats in GenAI systems, there is far less systematic support for assessing their privacy implications, leaving privacy analysis for GenAI-based applications comparatively underdeveloped.
To address this limitation, this paper presents a framework to support the privacy threat modeling of GenAI applications.

The new framework, built on top of \linddun{}, is constructed through a two-pronged approach: (i) an extensive analysis of the SotA in top conferences on privacy and security threats in GenAI, and (ii) a case-driven analysis of a GenAI-based application to identify the limitations of the existing privacy threat modeling framework and formulate new GenAI-specific privacy threat characteristics.
The resulting framework is subsequently applied to and evaluated over a different GenAI application case which involves an agentic AI assistant.
The outcome of these efforts is threefold. We provide (i) two detailed case studies of a privacy threat analysis of GenAI-based applications, (ii) a synthesis of the SotA privacy threats in GenAI applications, and (iii) a domain-specific GenAI privacy threat modeling framework, built on top of \linddun{}, that allows developers to leverage this privacy threat knowledge and analyze their GenAI applications.

The GenAI privacy threat modeling framework is designed to evolve. All threat knowledge sources, meta-model updates, and tooling are publicly available~\cite{supplementaryMaterials}, allowing the framework to be continuously updated as new privacy threats emerge in the scientific literature or in practice.

\section*{Data Availability Statement}
The supporting materials~\cite{supplementaryMaterials} to this work can be made available upon request.

\section*{Ethics Statement}

GenAI-based tools were used to revise the text, improve flow and correct any typos, grammatical errors, and awkward phrasing. All participants to the case studies are directly involved in this study, no external participants were involved. No additional ethical considerations apply. 

\section*{Acknowledgments}

The authors thank Tim Van hamme and Thomas Vissers for their assistance with the SotA synthesis.
\\ \\
This research was partially funded by the Research Fund KU Leuven, Internal Funds KU Leuven, and by the Cybersecurity Research Program Flanders.

\bibliographystyle{plain}
\bibliography{references.bib}

\appendix

\section{SotA Search Protocol}\label{app:search-protocol}
 \begin{figure*}
  \centering
     \includegraphics[width=.80\linewidth]{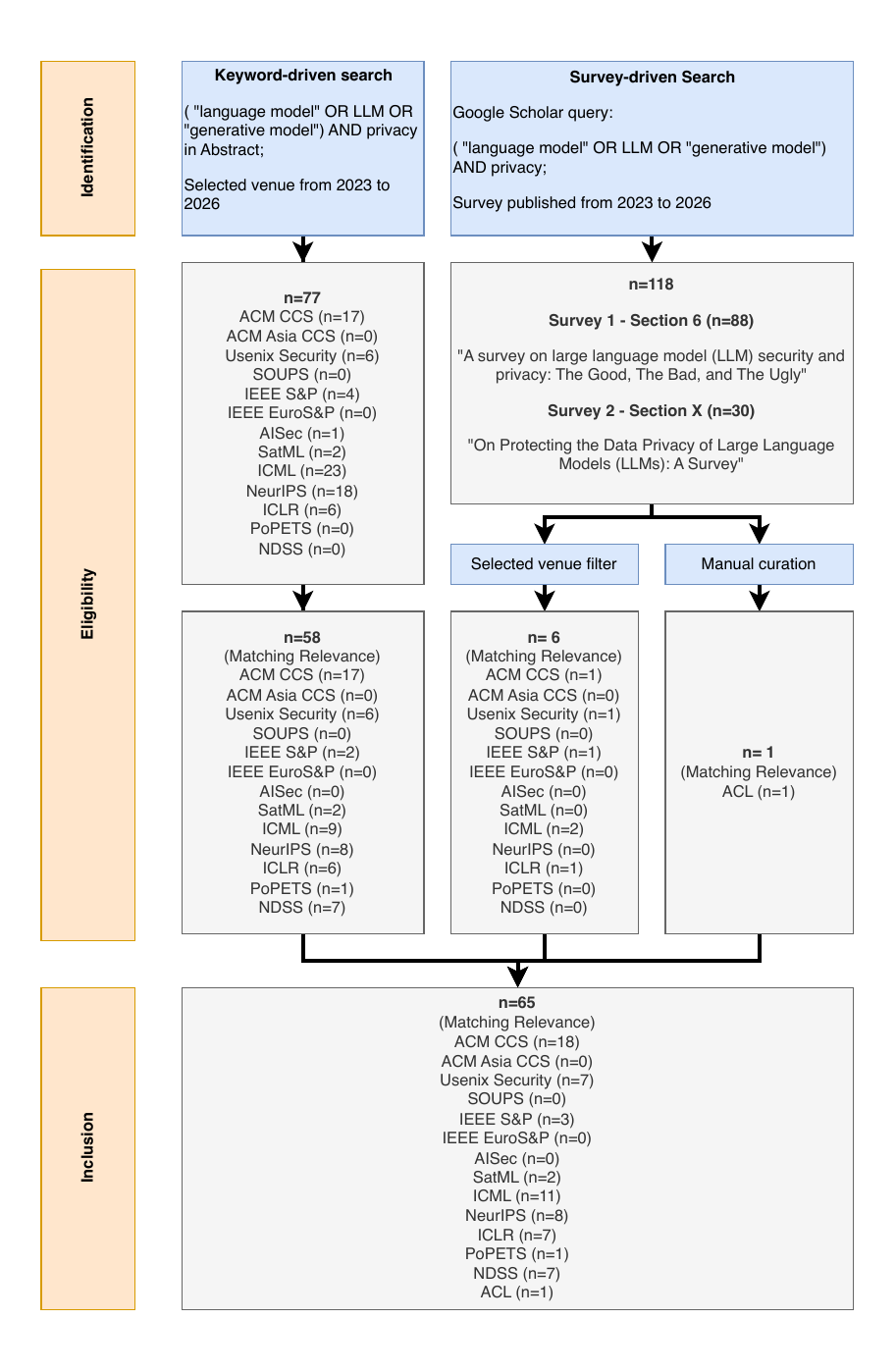}
     \caption{\label{fig:review}PRISMA diagram outlining the search protocol}
 \end{figure*}

\begin{lstfloat}\centering
\begin{minipage}{0.8\linewidth}
\begin{tcolorbox}[arc=0mm]
{\textbf{Venue list}}
        \begin{itemize}[nolistsep]
    \item ACM CCS
    \item ACM AsiaCCS
    \item AISec
    \item SatML
    \item USENIX Security
    \item IEEE S\&P
    \item IEEE EuroS\&P
    \item ICML
    \item NeurIPS
    \item ICLR
    \item PoPETS
    \item SOUPS
    \item NDSS
\end{itemize}
\end{tcolorbox}
\end{minipage}
\caption{\label{lst:searchvenues}Venue-driven Search}
\end{lstfloat}

\paragraph{First round: Keyword search} We searched top-tier academic venues (listed in~\cref{lst:searchvenues}) from 2023 to 2026 for papers whose abstracts contained both the term ``privacy'' and at least one of the following keywords: ``language model'', ``generative model'', or ``LLM''.
This ensures high-quality, peer-reviewed works that focus on novel threats and defenses for GenAI-empowered applications, rather than earlier proof-of-concept research or classic adversarial machine learning works.
Many papers in this phase addressed privacy threats using differential privacy and federated learning.
However, some recent attacks may have been missed if they used terms like ``data exfiltration'' or ``AI agents''. 
For the inclusion step, each of these papers is screened for relevance to this GenAI privacy study by manually evaluating the abstract.

\paragraph{Second round: survey-driven search}
To capture additional threats and mitigations published in those venues, we examined recent top surveys on GenAI privacy vulnerabilities and reviewed their specific threat and countermeasure sections. We located the official publications for these references and filtered them down to our list of primary venues, omitting duplicates from Phase 1. 
We removed the date restriction in this phase to include older—but highly relevant—surveyed topics that may not explicitly use our Round 1 keywords. 
For inclusion, each paper is screened for relevance to this GenAI privacy study by manually evaluating its abstract.

\paragraph{Third round: manual curation.}
We conducted a final manual review to include additional high-impact works on GenAI privacy threats, focusing on top-cited and highly relevant papers that might not appear in our primary venues. 
This ensures we capture influential research that could have been overlooked by earlier keyword or venue restrictions.

\paragraph{Inclusion and Exclusion Criteria.}
We include only papers that investigate privacy vulnerabilities and attacks in GenAI systems across data modalities, such as speech, images, and text.
To maintain a focused scope on privacy risks and evaluation in Generative AI systems, we excluded the following categories of work:

\begin{itemize}
    \item Well-established data modalities that have already been extensively studied in the privacy literature (e.g., synthetic tabular data generation).
    
    \item Purely cryptographic papers that focus on cryptographic primitives or protocol design without system-level privacy threat analysis.
    
    \item Security-focused attack papers that do not primarily address privacy risks, including poisoning attacks, adversarial attacks, model stealing attacks, and remote code execution exploits.
    
    \item Research on machine unlearning that does not explicitly analyze privacy threat models.
    
    \item Benchmarking or evaluation papers that do not explicitly discuss privacy implications.
    
    \item Works that enhance privacy using LLMs (e.g., employing LLMs as privacy judges or evaluators) without analyzing privacy threats introduced by GenAI systems.
    
    \item Studies where GenAI is positioned primarily as a privacy attacker, privacy protector, or privacy auditor, rather than as the system under privacy threat analysis.
\end{itemize}

\paragraph{Systematization of knowledge.}
In total, 65 papers are described and classified. 
We conducted a thorough review of the literature based on the selected papers. We identified the privacy risk scenarios covered in these works and mapped them onto the \linddun{} framework.
\balance
In curating the literature for our synthesis, we further exclude works whose primary contribution is the design of defense mechanisms across multiple threat models, as these papers focus on mitigation strategies rather than on specifying the underlying threat assumptions. 
Similarly, we exclude papers centered on security backdoors or model compromise, since these fall under the domain of system security rather than privacy and do not align with our objective of analyzing privacy-relevant threat models. 
Our goal is therefore to restrict the corpus to papers that explicitly articulate privacy threat models or attack assumptions, ensuring conceptual clarity and consistency in the mapping to \linddun{} threat types.

\end{document}